\documentclass[sigplan,screen]{acmart}

\AtBeginDocument{%
  \providecommand\BibTeX{{%
    \normalfont B\kern-0.5em{\scshape i\kern-0.25em b}\kern-0.8em\TeX}}}

\setcopyright{rightsretained}
\emergencystretch=\maxdimen
\makeatletter
\def\@copyrightspace{\relax}
\makeatother
\begin{document}

\title{Mapping Computer Science Research: Trends, Influences, and Predictions}


\author{Mohammed Almutairi}
\affiliation{%
  \institution{University of Notre Dame}
  \city{Notre Dame}
  \country{USA}}
\email{malmutai@nd.edu}

\author{Ozioma Collins Oguine}
\affiliation{%
  \institution{University of Notre Dame}
  \city{Notre Dame}
  \country{USA}}
  \email{ooguine@nd.edu}

\renewcommand{\shortauthors}{M. Almutairi and O.C. Oguine}

\begin{abstract}
This paper explores the current trending research areas in the field of Computer Science (CS) and investigates the factors contributing to their emergence. Leveraging a comprehensive dataset comprising papers, citations, and funding information, we employ advanced machine learning techniques, including Decision Tree and Logistic Regression models, to predict trending research areas. Our analysis reveals that the number of references cited in research papers (Reference Count) plays a pivotal role in determining trending research areas making reference counts the most relevant factor that drives trend in the CS field. Additionally, the influence of NSF grants and patents on trending topics has increased over time. The Logistic Regression model outperforms the Decision Tree model in predicting trends, exhibiting higher accuracy, precision, recall, and F1 score. By surpassing a random guess baseline, our data-driven approach demonstrates higher accuracy and efficacy in identifying trending research areas. The results offer valuable insights into the trending research areas, providing researchers and institutions with a data-driven foundation for decision-making and future research direction.

\end{abstract}




\maketitle
\pagestyle{plain}

\section{Introduction}
The rapidly evolving landscape of research ideas and advancements in computing power has sparked a revolution in the field of Computer Science (CS). These revolutionary changes have brought about a shift in research trends, influencing crucial aspects such as funding, research interests, and career opportunities within the CS domain. While researchers have made commendable efforts to comprehend the polarization of research fields in Computer Science through data analysis of factors like citation scores \cite{11}, top active authors ~\cite{15}, and funding \cite{11}, there are still gaps in our comprehensive understanding of these trends. Our research bridges these gaps by expanding the current state-of-the-art knowledge and gaining a broader understanding of the diverse trends within the domain of Computer Science.

According to a report by the U.S. Bureau of Labor Statistics \cite{10}, occupations in computer and information technology are projected to experience an 11\% growth between 2019 and 2029. Additionally, a report \cite{9} estimates the presence of over 1.35 million tech startups globally as of 2021, leading to increased tech investment and global research activities. While these statistics highlight the existence of polarized interests within the computer science and information technology domain, there is still much to uncover regarding the reasons behind these emerging trends, the factors influencing them, and the ability to predict future trends accurately.

The advent of big data \cite{20} and data science techniques has facilitated the efficient tracking of trends over different time periods. However, understanding these trends requires a nuanced approach as multiple studies, built on diverse hypotheses, have provided varying perspectives and findings on the evolution of trends within the Computer Science domain. For example, a study conducted by Hoonlor et al. \cite{11} explored research communities, trends, and the relationship between awarded grants and changes in communities and trends. Their research revealed a correlation between the frequency of publication topics and an increase in funding. Furthermore, their findings demonstrated the tendency of research fields within CS to address new challenges using state-of-the-art technologies. Another paper by Effendy and Yap utilized the Field of Study (FoS) classification technique to analyze trends in the Computer Science domain based on citations.

Building upon these previous research findings, our study provides a holistic view of the changes that have transpired within research fields in the domain of computer science over time. To achieve this, we formulated the following research questions:\\ 
\textbf{RQ1:} What are the current trending research Areas in the field of computer science? \\
\textbf{RQ2:} What factors contribute to the emergence of these trending research areas in Computer Science? \\
\textbf{RQ3:} How do these trends evolve over time? \\

To answer these research questions, we first conducted data preparation and preprocessing, ensuring data consistency and handling noisy and missing data. Featurization was performed to select relevant features and create new ones, including the 'Citation Age,' which helped address citation outliers. We then proceeded with model selection, choosing the Decision Tree and Logistic Regression models for their predictive capabilities. After model training and testing, our findings revealed that the number of references cited in research papers (Reference Count) significantly influences trending research areas. Additionally, the impact of NSF grants and patents on trends increased over time. The Logistic Regression model outperformed the Decision Tree model, exhibiting higher accuracy, precision, recall, and F1 score. These results provide valuable insights into the evolving landscape of computer science research and offer a data-driven foundation for researchers and institutions to make informed decisions and plan future research endeavors.
\section{Methods}

\subsection{Dataset}
We utilized the Microsoft Academic Graph (MAG) dataset ~\cite{microsoft_academic_2022_6511057} which captures the research of science in different fields ~\cite{12}. MAG is a heterogeneous dataset that has scientific publications which offers a comprehensive of the global research output. The dataset is created and maintained by Microsoft Research who provides it for freely academic use ~\cite{13}. For our project, we considered data collected between the time period of 2012 -2021.

\subsubsection{Featurization}
\begin{figure*}
    \centering
    \includegraphics[width=0.7\linewidth]{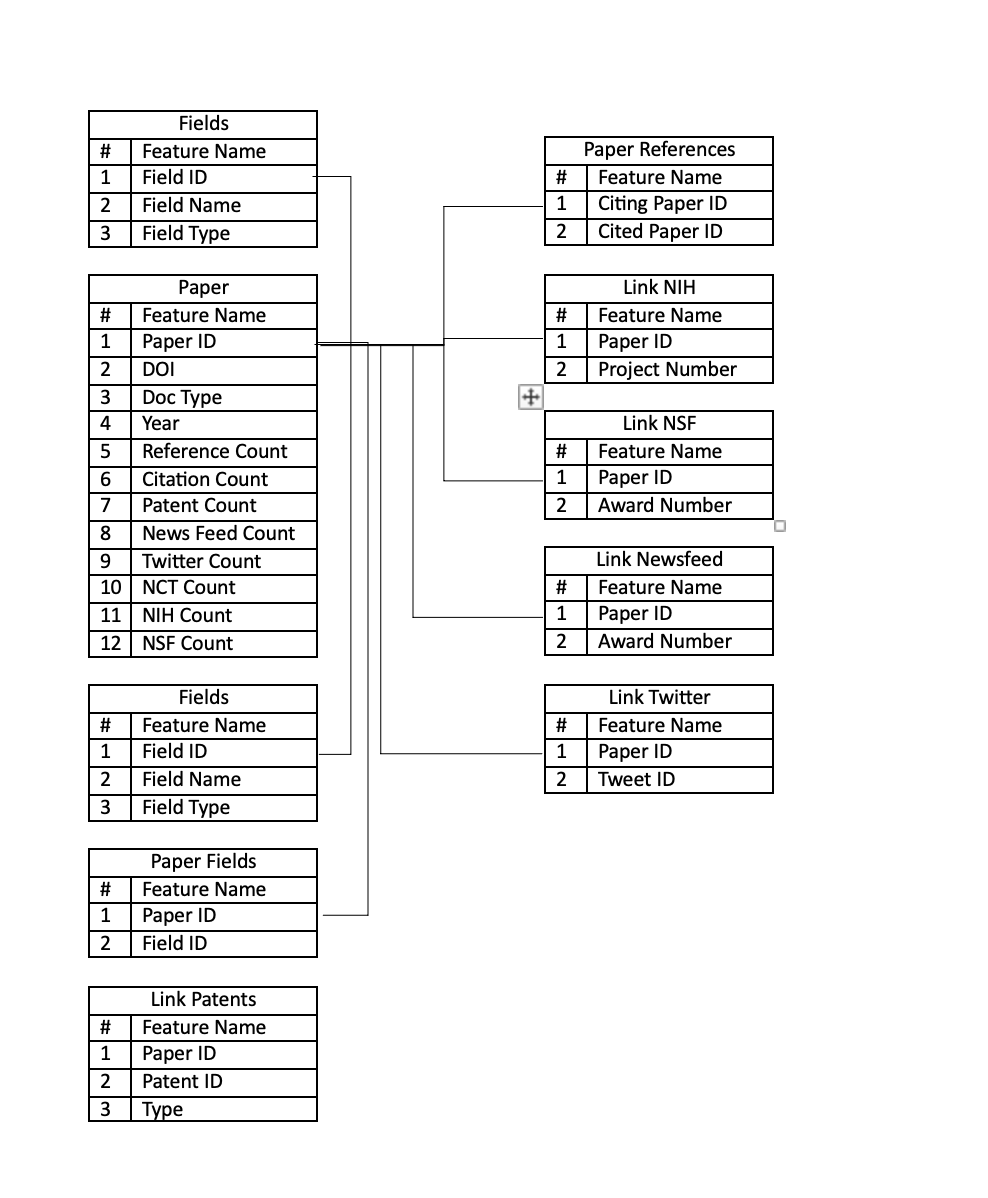}
    \caption{Features and Tables used to train and test the models}
    \label{1}
\end{figure*}

We selected a set of features in order to perform necessary data preprocessing. The features are 'Paper ID', 'DOI', 'Doc Type', 'Year', 'Journal ID', 'Conference Series ID', 'Citation Count', 'Reference Count', 'News feed Count', 'Tweet Count', 'NCT Count', 'NIH Count', and 'NSF Count'.  We created new features such as the ‘Citation Age’ which represents the average citation rate per year in order to smooth out the Skewness of the ‘Citation Count’ feature (See Figure \ref{1})

\subsection{Data Preparation and Preprocessing:}

The process of data preparation and preprocessing was a critical phase in our experimental process. In this section, we outline the various stages involved in preparing and preprocessing the dataset for our study. The dataset used for this research is a subset of the MAG (Microsoft Academic Graph) dataset, containing several sets such as Papers, Paper Details, Link\_NSF, Link NIH, Paper Fields, Fields, and Link Twitter. Integration of these subsets is performed using the Paper-ID (unique identifier) as the primary key to facilitate correct data linking.

\subsubsection{Ensuring Data Consistency:}
One of the initial challenges encountered in the data was inconsistency. For instance, some records had the year written as "03," while others represented the same year as "2003." To ensure data consistency, we standardized the format to represent all years in a uniform manner (yyyy), which is particularly crucial for accurate temporal and time series analysis.

\subsubsection{Handling Noisy Data:}
Another problem we encountered in our data preparation and preprocessing was the issue of noisy data. In our case, one form of noise was related to the Paper-ID field, where some papers had alphanumeric identifiers while others had numeric identifiers. Consequently, linking this field to other tables became problematic due to the presence of numeric IDs in those tables. To address this issue, we standardized the Paper-ID field by replacing the alphanumeric identifiers with numeric identifiers, which provided more consistency across different tables. By comparing the DOIs, we were able to establish links and unify the Paper-IDs, resolving the noise and ensuring proper data linkage.

\subsubsection{Dealing with Missing Data:}
We encountered two types of missing data in our dataset: Missing Completely at Random (MCAR) and Missing at Random (MAR). MCAR refers to data points that are missing randomly across the dataset, with no systematic pattern. For example, we discovered that paper types were missing which could be an error in data entering. In contrast, MAR implies that the missing data is dependent on other observed variables. For example, we observed that the DOI was missing for papers published before 19900. This constitutes a case of MAR, where the missingness is linked to the publication year. To handle these cases, we considered appropriate imputation of paper type with unknown categories while for the missing DOIs we deleted missing values.

\subsubsection{Dealing with Citation Outliers:}
\begin{figure}
    \centering
    \includegraphics[width=0.8\linewidth]{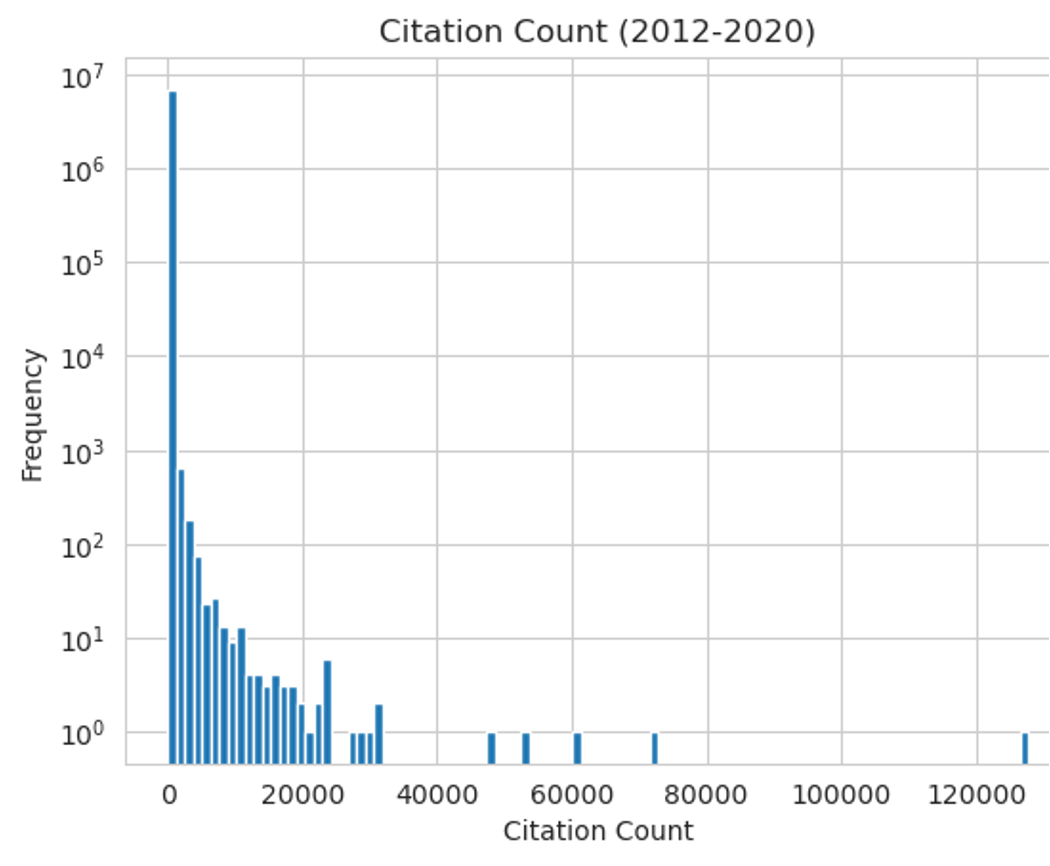}
    \caption{A figure shows that some papers received very few citations, while others garnered a significant number}
    \label{2}
\end{figure}

The citation feature in the dataset presented another challenge due to the inherent variability in citation counts. We discovered that some papers received very few citations, while others garnered a significant number which we uncovered using visual inspection (Figure \ref{2}). To address this, we created a new feature called "Citation Age" designed to capture the average citation rate per year since publication. This approach helps normalize the citation count by accounting for the number of years since a paper's publication. For instance, a paper published in 2015 with 100 citations by 2023 would have a Citation Age of 12.5 (100 citations divided by 8 years). This transformation helps mitigate the impact of extreme outliers, providing a more balanced representation of citation patterns across different papers.

\subsection{Data Analysis}
For our research, we employed data-driven approaches to analyze the dataset and answer our research questions. The data analysis processes are:

\subsubsection{Model Selection:}
In this phase, we considered different models based on the data type and the nature of our research questions. Our dataset consisted of both categorical and numerical data types. After careful consideration, we selected two models that best suited our needs:

\begin{itemize}
  \item Decision Tree Model: We picked Decision trees because they are well-suited for handling both categorical and numerical data, making them a suitable choice for our dataset. Additionally, decision trees provide interpretable results, allowing us to understand the factors that influence the prediction of paper trending.
  
\item Logistic Regression Model: We chose logistic regression as it is specifically designed for binary classification tasks. Given that our dataset involves predicting whether a paper is trending or not, the logistic regression model was an appropriate choice. Moreover, logistic regression provides coefficients that indicate the importance of each feature in predicting the outcome, offering valuable insights into the factors driving paper trending.
\end{itemize}
\subsubsection{Model Training and Testing:}
Since decision trees and logistic regression do not require extensive data preprocessing, we directly proceeded to model training and testing with our dataset. To comparatively evaluate the performance of the selected models, we divided the dataset into two splits: a training set and a test set. The training set comprised 80\% of the entire dataset and was used to train both the decision tree and logistic regression models while the test set (20\% of our dataset) was used to validate the generalizability of the model performance. We ensured that the testing set had not been utilized during the training process, enabling us to accurately assess the models' capabilities and avoiding overfitting.

\subsubsection{Model Evaluation Metrics}
To evaluate the performance of our models, we have chosen multiple metrics to assess their predictive capabilities. The following metrics was used:
\begin{itemize}
    \item Accuracy: Accuracy measures the overall correctness of our model's predictions. It is calculated as the ratio of correctly predicted instances (both trending and not trending) to the total number of instances in the dataset. A higher accuracy value indicates a better-performing model.

    \item Precision: Precision is the proportion of true positive predictions (trending papers correctly identified as trending) to the total number of predicted positive instances (papers predicted as trending). It measures our model's ability to correctly identify true positives and avoid false positives.

    \item Recall (Sensitivity): Recall, also known as sensitivity, is the proportion of true positive predictions to the total number of actual positive instances (papers that are actually trending). It represents our model's ability to capture all positive instances correctly.

    \item F1 Score: The F1 score is the harmonic mean of precision and recall. It provides a balanced measure between precision and recall, especially when dealing with imbalanced datasets where one class dominates the other. A high F1 score indicates a good balance between precision and recall.
 
\end{itemize}

\begin{table}[ht]
  \caption{Performance Evaluation of Decision Tree Model}
  \label{tab:1}
  \begin{tabular}{lrr}
    \toprule
  Trending&Decision Tree&Logistic Regression\\
    \midrule
    Accuracy & 0.73& 0.78  \\
    Precision & 0.91& 0.89  \\
    Recall & 0.72 & 0.81 \\
    F1-score &  0.81 & 0.85\\
  \bottomrule
\end{tabular}
\end{table}

In Table~\ref{tab:1}, we present the results of the comparative evaluation of two predictive models: Decision Tree and Logistic Regression. These metrics provide insights into the performance of each model in predicting the classes correctly.

The Decision Tree model achieved an accuracy of 0.73, which means that it correctly predicted approximately 73\% of the total instances. The precision of the Decision Tree model was 0.91, indicating that out of the instances predicted as positive (trending), 91\% were true positive predictions. The recall for the Decision Tree model was 0.72, showing that it captured 72\% of the actual positive instances (trending papers) correctly. The F1 Score, which considers both precision and recall, was 0.81 for the Decision Tree model.

On the other hand, the Logistic Regression model outperformed the Decision Tree model in terms of accuracy, recall, and F1 Score. The Logistic Regression model achieved an accuracy of 0.78, indicating a higher overall correctness in its predictions. It also had a precision of 0.89, meaning that 89\% of the predicted positive instances were true positive predictions. The recall for the Logistic Regression model was 0.81, indicating a higher ability to correctly identify actual positive instances. The F1 Score for the Logistic Regression model was 0.85, reflecting a more balanced performance between precision and recall.

\begin{figure}{h}
    \centering
    \includegraphics[width=0.8\linewidth]{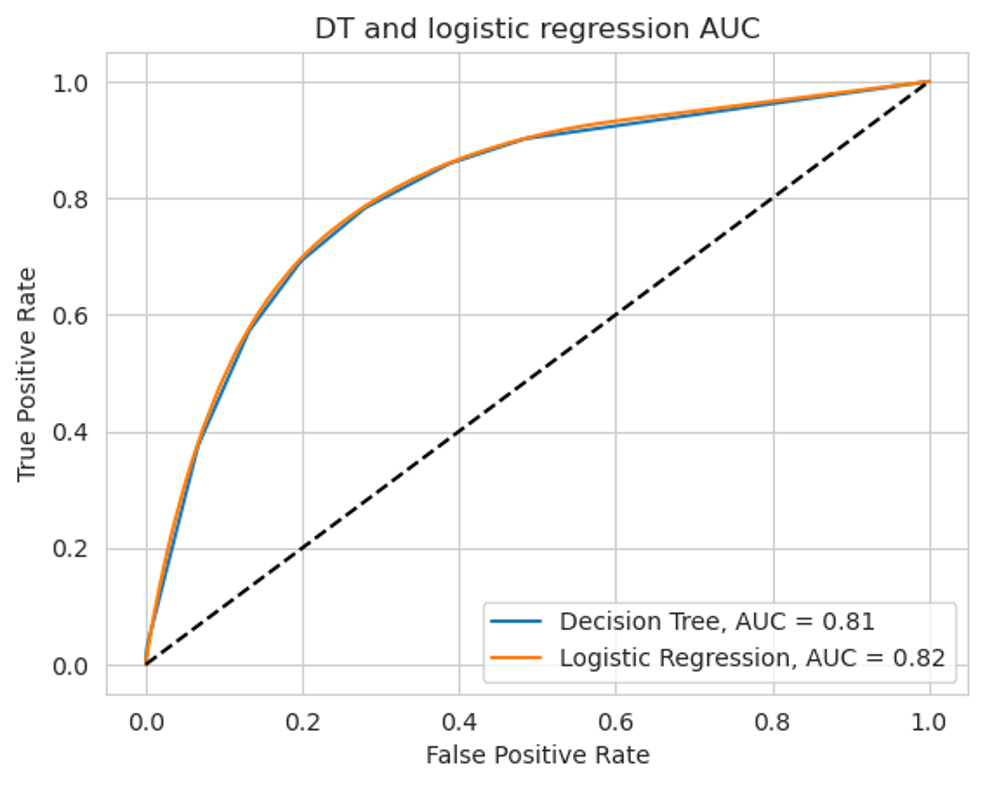}
    \caption{ROC Curve for Decision Tree and Logistic Regression}
    \label{3}
\end{figure}

Based on the comparative evaluation, the Logistic Regression model appears to be the better-performing model, as it exhibits higher values for accuracy, recall, and F1 Score compared to the Decision Tree model. To verify the claim and further assess the performance of the two models (Decision Tree and Logistic Regression) for our binary classification (Trending or not Trending), we plotted a Receiver Operating Characteristic (ROC) curve shown in Figure \ref{3} where we observe an AUC value of 0.82 for Logistic Regression and an AUC value 0.81 for Decision tree.  
  
\section{Result and Discussion:}
\subsection{What are the current trending research Areas in the field of computer science?}
\begin{figure}
    \centering
    \includegraphics[width=0.8\linewidth]{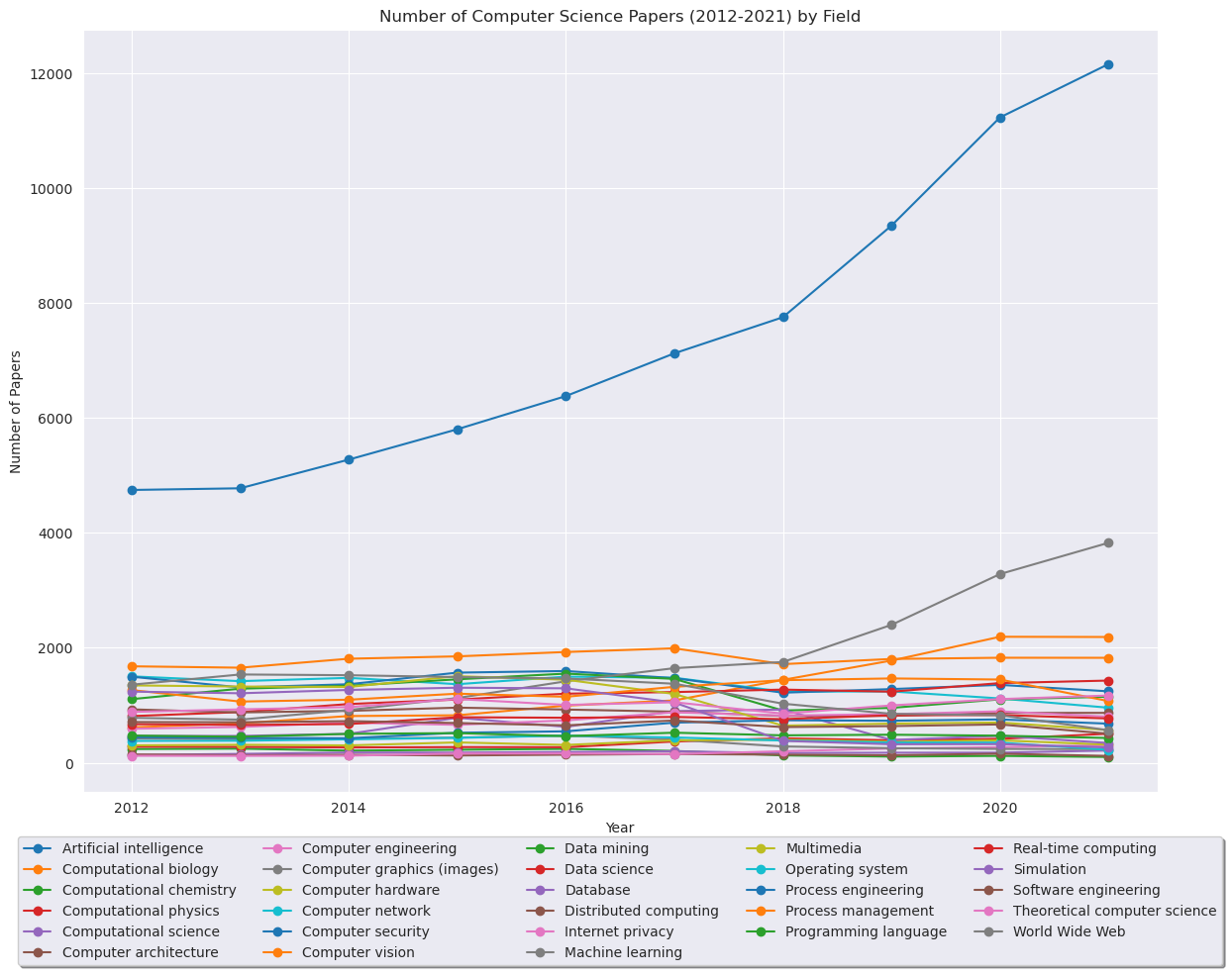}
    \caption{Number of Computer Science Papers (2012-2021) by Fields}
    \label{4}
\end{figure}

\begin{figure} 
\centering
    \includegraphics[width=0.8\linewidth]{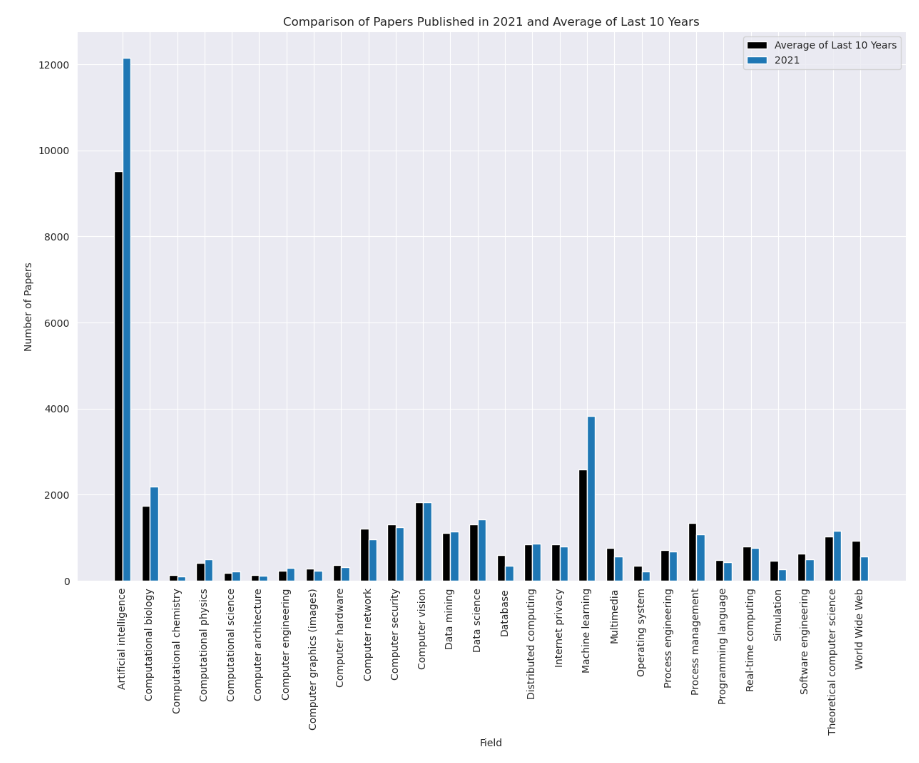}
    \caption{Comparison of Papers published in 2021 and Average of the Last 10 years}
    \label{5}
\end{figure}

To identify the trending research areas in computer science, we utilized two measures. Firstly, we analyzed the number of papers published within each year. By examining the publication trends over time, we identified areas that have seen a significant increase in research activity and publication volume, indicating their current popularity and relevance in the field as shown in Figure \ref{4} and \ref{5}.

\subsection{What factors contribute to the emergence of these trending research areas in Computer Science?}
For RQ2, we employed two predictive models, the Decision Tree model and the Logistic Regression model, to identify factors contributing to the emergence of trending research areas in computer science.

\begin{figure}
    \centering
    \includegraphics[width=0.8\linewidth]{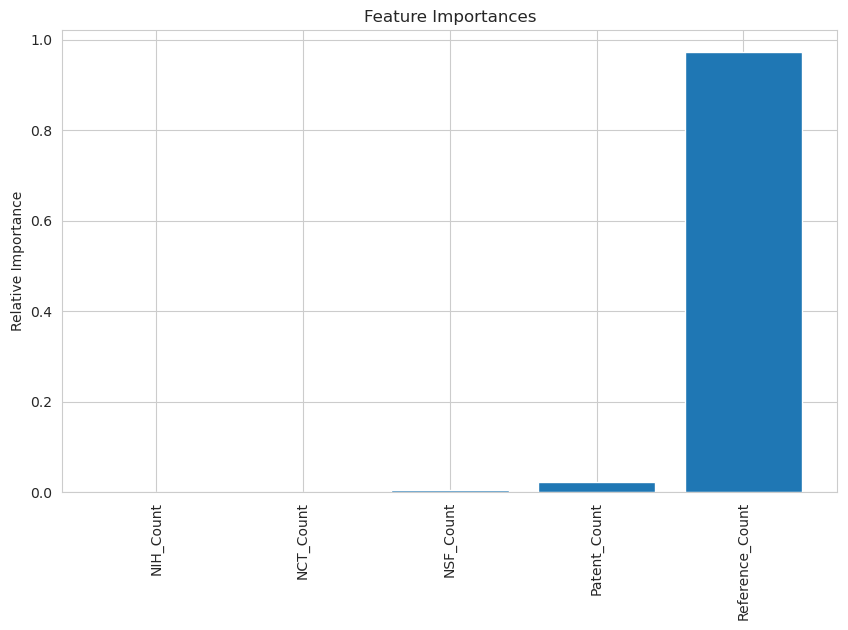}
    \caption{Bar chart showing the feature importance from the Decision Tree }
    \label{6}
\end{figure}

According to the Decision Tree model, the Reference Count emerged as an important predictor in determining trending research topics (shown in \ref{6}). This suggests that the number of references cited within a paper plays a crucial role in the popularity and emergence of specific research areas. Papers with higher reference counts may indicate topics that are gaining significant attention and interest from researchers, leading to their classification as trending research areas.

\begin{figure} 
\centering
    \includegraphics[width=0.8\linewidth]{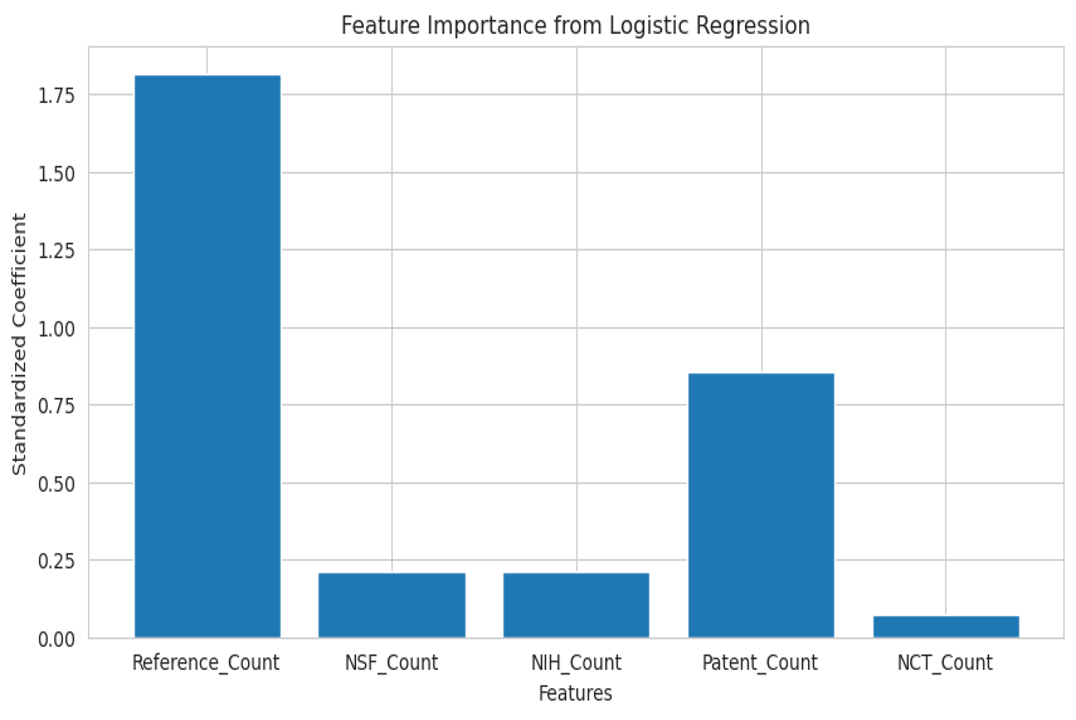}
    \caption{Bar chart showing the feature importance from the Logistic Regression}
    \label{7}
\end{figure}

On the other hand, the Logistic Regression model also highlights the Reference Count as the most influential predictor of a topic trending (shown in \ref{7}). This supports the findings from the Decision Tree model, reinforcing the significance of reference counts in determining the popularity of research areas in computer science.

\subsection{How do these trends evolve over time?}
To understand how the trends in computer science research fields evolve over time, we divided the analysis into three periods: the first period (2012-2015), the second period (2015-2019), and the third period (2020-2022) and observed the importance of different features in predicting trending research fields during each period.

During the first period (2012-2015), the most important feature for predicting trending research areas was the Reference Count. Additionally, the Patent Count also held some importance in this period, suggesting that patents played a role in shaping the emerging research fields at the time.

Moving to the second period (2015-2019), the Reference Count remained the most important feature for predicting trending research areas. However, we observed a slight increase in the importance of NSF Count and Patent Count during this period. This indicates that the influence of NSF grants and patents in contributing to trending research fields grew in comparison to the previous period.

In the third period (2020-2021), we found that the importance of the Reference Count continued to dominate as the most significant predictor of trending research fields. Furthermore, the importance of the Reference Count further increased compared to the previous periods.

\begin{figure}
    \centering
    \includegraphics[width=0.8\linewidth]{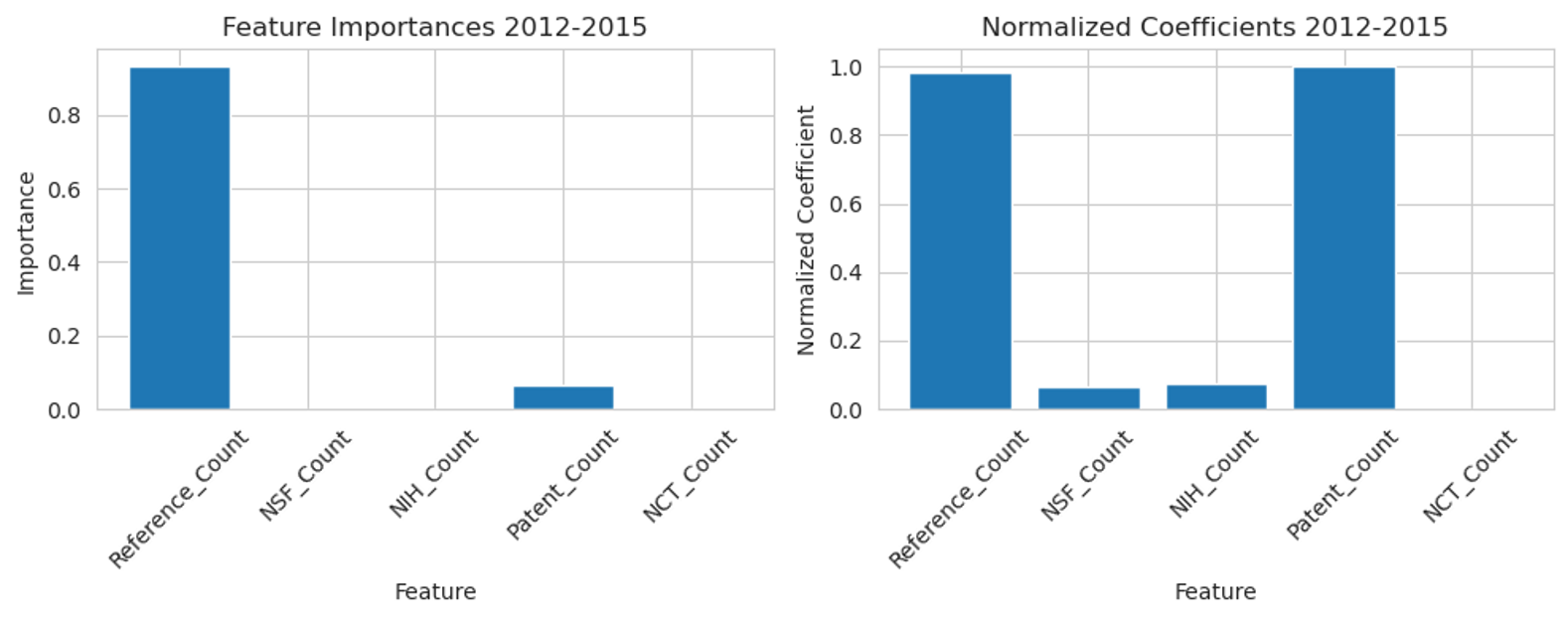}
    \includegraphics[width=0.8\linewidth]{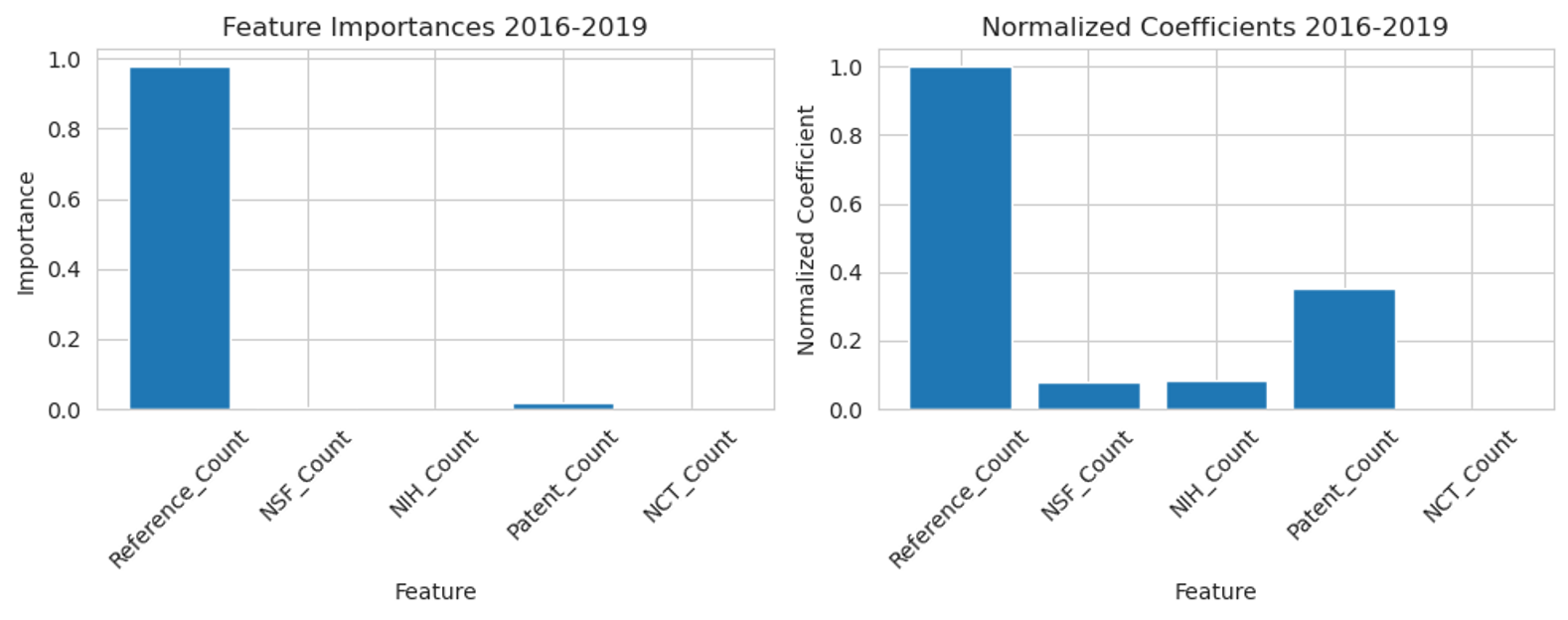}
    \includegraphics[width=0.8\linewidth]{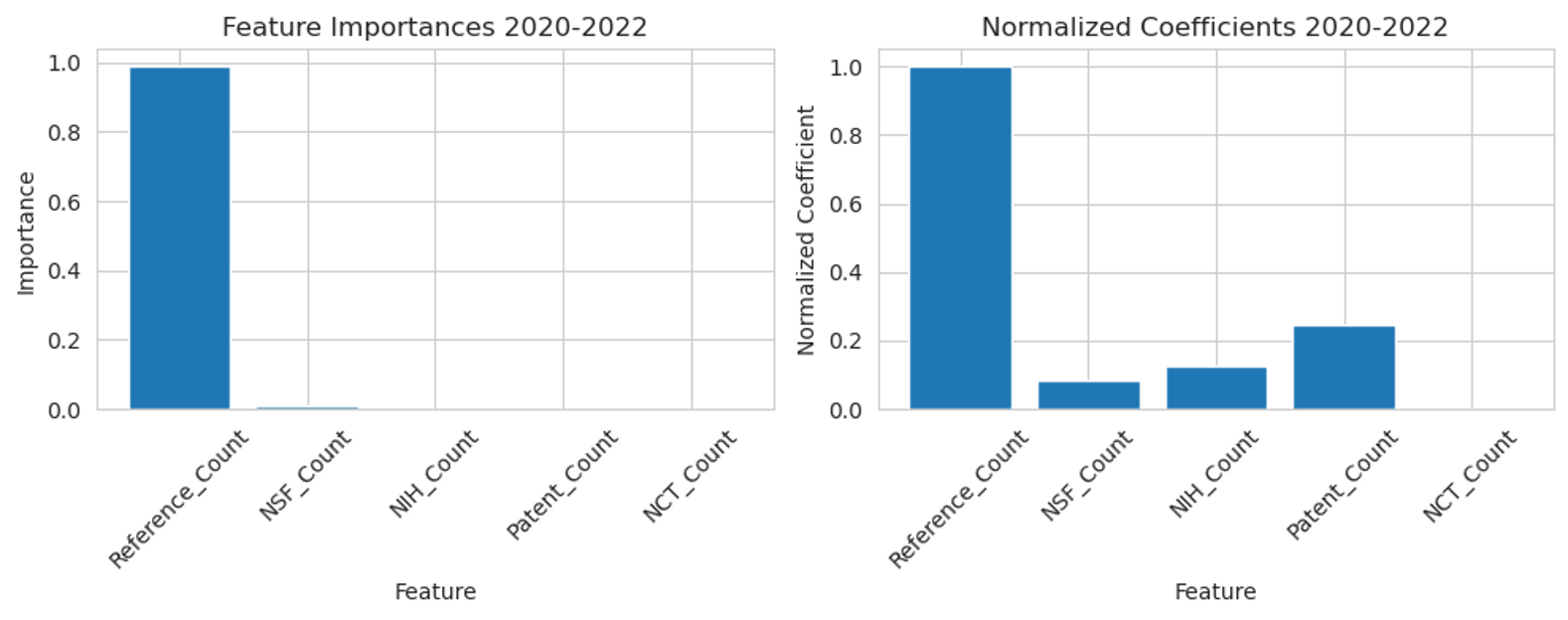}
    \caption{Factors Importance Ranking across three time periods}
    \label{8}
\end{figure}

Overall, the evolving trend across all three periods (as shown in Figure \ref{8})indicates that the Reference Count consistently plays a crucial role in shaping trending research areas in computer science. As the number of references cited in research papers increases, it signifies the impact and relevance of the work, driving the popularity of certain research topics.

\subsection{Baseline Comparison}
As a baseline comparison, we established a random guess baseline with an accuracy of 49.8\%. This means that a random prediction would correctly classify approximately half of the instances in the dataset. By contrast, our predictive models, such as the Decision Tree and Logistic Regression, achieved higher accuracies, indicating their superior performance in analyzing trends in computer science research areas. Our models success in outperforming the baseline underscores its value in providing a more accurate classification of computer science research trends.

\section{Summary and Conclusion:}
In our study, we aimed to analyze the current trending research areas in the field of computer science and understand the factors contributing to their emergence. To achieve this, we utilized a data-driven approach, analyzing a comprehensive dataset containing information on papers, their citations, and funding. We employed various data preprocessing techniques and feature engineering to derive meaningful insights from the data.

Our approach involved using two predictive models, the Decision Tree model and the Logistic Regression model, to identify key factors influencing the trends in computer science research areas. The results of our analysis provided unique insights into the dynamics of trending topics in computer science and their evolution over time.

The findings from our study reveal that the number of references cited in research papers, represented by the Reference Count, plays a central role in determining trending research areas. As the Reference Count increases, it signifies the significance and impact of research works, leading to the popularity of specific research topics. Moreover, we observed a growing influence of NSF grants and patents on trending research areas over time, indicating their increasing importance in shaping the computer science research landscape.

The utilization of multiple evaluation metrics, including Accuracy, Precision, Recall, and F1 Score, allowed us to assess the performance of our models thoroughly. The models demonstrated commendable accuracy in predicting trending research areas, with the Logistic Regression model outperforming the Decision Tree model.

\section{Limitation}
We acknowledge that the feature vectors used in our research were chosen subjectively, which may limit the generalizability of our results. Furthermore, our research findings were influenced by various constraints, including limited computational resources, time constraints, and dataset complexity. Therefore, future studies in this area should make a concerted effort to consider these factors carefully to achieve more reliable and improved results.

\begin{acks}
We want to acknowledge Microsoft for making available the Microsoft Academic Graph (MAG) dataset.
\end{acks}

\bibliographystyle{ACM-Reference-Format}
\bibliography{sample-base}

\appendix
\end{document}